***Preprint: Accepted to be published in ELSEVIER Book Series SUSTAINABLE DIGITAL MEDICINE: ISBN: 9780443458606***

# Designing Technology for Social Wellbeing in Built Environments: A Conceptual Framework

Gul Sher Ali[1*], Michail Giannakos[1], Monica Lillefjell[2] and Sobah Abbas Petersen[1]

[1]Department of Computer Science
[2]Department of Neuromedicine and Movement Science
Norwegian University of Science and Technology, Trondheim, Norway
gulsher.ali@ntnu.no[*] , michailg@ntnu.no , monica.lillefjell@ntnu.no , sap@ntnu.no

**Abstract:** Digital technologies are deployed in urban built environments with the aim of supporting social dimensions. However, the research offers no unified guidance for such digital technology design. Additionally, evidence shows that week social wellbeing contributes to mental and physical health outcomes which suggests that the technology designed to strengthen social wellbeing could also function as a form of health promoting and preventive intervention. This chapter addresses this research gap by developing a conceptual framework that supports the design of digital technologies for social wellbeing in built environments. We propose a conceptual framework composed of three core components which are drawn on the synthesis of selected empirical studies on technologies embedded in built environments for social wellbeing. First, a social wellbeing dimensions model that identifies what digital technology could address. Second, a digital technology contribution matrix that distinguishes the types of contributions a digital technology could make. Third, levels that maps the scope at which technology could support social wellbeing. This conceptual framework could help researchers, practitioners and policymakers to design and guide digital technology interventions that target social wellbeing in the built environment.

## 1. Introduction

The built environment, including buildings, public spaces, and infrastructure, plays a central role in shaping human health and wellbeing (Andalib et al., 2024; Lach et al., 2022). The health and wellbeing outcomes are multidimensional (Andalib et al., 2024) such as a physical wellbeing which depends on indoor air quality, thermal comfort, and access to green space. The mental wellbeing is influenced by noise, crowding and restorative environments The social wellbeing is shaped by how people meet, form communities and develop a sense of belonging in the physical settings where they live, work and interact (Keyes, 1998).

Considering this importance, the research emphasized using digital technologies within built environments to address one or more of these wellbeing dimensions. Such as, smart building systems target physical comfort and energy efficiency, digital health tools monitor physiological parameters and place-based technologies aim to support social connection (Marshall et al., 2024; Nikolic and Yang, 2020; Rainisio et al., 2024). Yet the social dimension of wellbeing remains comparatively neglected in digital technology design (Han and Kim, 2024; Yang and Kim, 2024).

Social wellbeing refers to the quality of an individual's relationships, community engagement, and sense of belonging (Keyes, 1998). The social wellbeing is distinctive because it operates at the interface between individuals and their communities and mediate how people experience and benefit from their environments. Social wellbeing is not independent of other dimensions but interconnected with them, i.e., strong social ties promote mental health by buffering stress and reducing loneliness (Hodgson et al., 2020; Holt-Lunstad, 2024), social participation encourages physical activity and healthier behaviours (Wang et al., 2023), and equitable social inclusion determines who benefits from environmental improvements (Keyes, 1998). On the other hand, when social wellbeing deteriorates, the effects cascade as the research shows that the social isolation is associated with increase in all-cause mortality risk, comparable to established risk factors such as smoking and physical inactivity (Holt-Lunstad, 2024; Wang et al., 2023). Similarly, the World Health Organization (WHO) has called for social aspects to be treated with the same urgency as physical and mental aspects (Garcia et al., 2025). This positions social wellbeing not merely as a quality-of-life outcome but as a pathway through which built environment digital interventions (Marshall et al., 2024; Nikolic and Yang, 2020; Rainisio et al., 2024) can influence broader health outcomes. More specifically, the digital technologies that strengthen social interaction, community connection, place attachment and equitable participation in built environments can therefore function as a form of health promoting and preventing intervention.

Despite this much importance of digital technologies in built environments for social wellbeing, the designers of such technologies have little structured guidance for addressing it comprehensively. Existing studies tend to focus on isolated aspects such as social interaction (Hatem et al., 2024; Nikolic and Yang, 2020), sense of place (Colangelo et al., 2023; Sharji et al., 2022), community satisfaction (Rainisio et al., 2024), or social inclusion (Marshall et al., 2024), without providing an integrated foundation that technology designers could use to ensure that their digital interventions address social wellbeing from all possible perspectives.

As a result, the field lacks a way to identify what social wellbeing dimensions a digital technology could target, what types of contributions it could aim for, and at what scope it could operate. There are existing frameworks for wellbeing in built environments that address parts of this challenge but leave the digital technology mediated social dimension largely unexamined. One recent framework (Makaremi et al., 2025) integrates individual, collective and environmental elements of wellbeing in the built environment but focuses on building-level evaluation rather than technology-specific design guidance. Another (Tan et al., 2025) develops a residential wellbeing model but does not address how technology interventions could be designed to shape social outcomes. A third (Akbarinejad et al., 2023) proposes a social sustainability assessment for built environments but treats technology as a contextual factor rather than a designable intervention. There is no framework that connects digital technology design decisions to social wellbeing outcomes in a way that is both theoretically grounded and practically usable.

This chapter addresses this research gap by developing a conceptual framework for designing digital technology that promotes social wellbeing in built environments. We draw on empirical studies selected because they designed and examined technologies in a built environment and report findings relevant to social wellbeing. We synthesize this evidence using a conceptual framework analysis approach (Jabareen, 2009) which resulted in a conceptual framework with three components. The first component is social wellbeing dimensions identifying what digital technologies could address and how social wellbeing dimensions interrelate. The second component is a digital technology contribution matrix distinguishing the types of contributions technology could make along procedural-substantive and subjective-objective axes, and finally, the scope mapping the levels at which digital technology could support social wellbeing from individual-level processes to systematic integration. This conceptual framework together with operationalization guidelines provide a structured tool for designing digital technologies that target social wellbeing and for guiding policy on technology integration in the built environment. It also contributes to the broader agenda of sustainable health measures by grounding digital technology design in the social determinants of health (Dahlgren and Whitehead, 2021).

The remainder of this chapter is organised as follows. Section 2 presents the review of the empirical studies on technology for social wellbeing in built environment and analytical approach that grounds our conceptual framework. Section 3 introduces the conceptual framework and its three components. Section 4 integrates these components into practical use through measurement and design guidance and a usage protocol. Section 5 discusses the framework's contribution in relation to existing work and considers its implications for researchers, practitioners, and policymakers, along with its limitations. Finally, the section 6 concludes with directions for future research.

## 2. Literature and Analytical Approach

This section reviews the empirical studies that examine technological interventions in built environments designed to support social wellbeing outcomes. We draw on a focused body of empirical work reported in the literature, selected because each study examines a technology intervention in a built environment with relevance to social wellbeing goals. While some of these technologies were designed with broad objectives (e.g., environmental monitoring or healthcare delivery), they all incorporated features intended to support social outcomes, making them appropriate cases for extracting design-relevant knowledge about what works, what fails and why. Although the corpus spans settings from public spaces and healthcare to residential and recreational environments, our analytical focus is on the social wellbeing outcomes rather than the built environment typologies described in the studies. This methodology is illustrated in Figure 1.

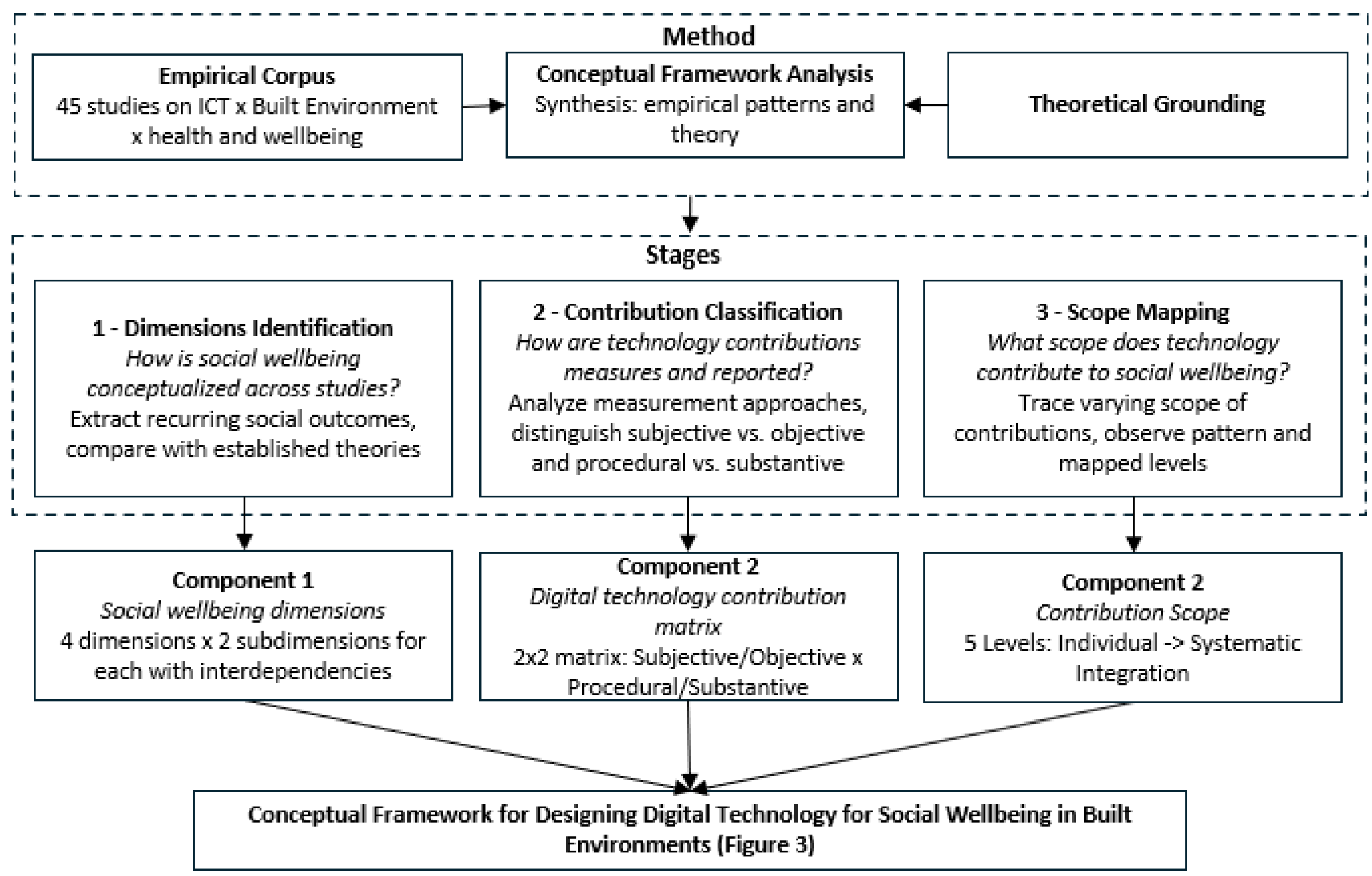

Figure 1: Methodological Process: Synthesising Empirical Evidence into a Conceptual Framework

To move from this empirical evidence toward a conceptual framework, we adopt a conceptual framework analysis approach (Jabareen, 2009), which involves synthesizing existing empirical evidence to construct a new conceptual framework rather than testing hypotheses or extracting variables through systematic coding. The analysis proceeds in three stages. First, we examine how social wellbeing is conceptualized across studies, identifying recurring dimensions and their subdimensions through comparison with established theoretical accounts, particularly Keyes's multidimensional model of social wellbeing (Keyes, 1998). This produced the social wellbeing dimensions model (Component 1). Second, we analyse the measurement approaches and outcome types reported across studies, distinguishing between subjective and objective measures and between procedural and substantive contributions which formed the digital technology contribution matrix (Component 2). Third, we trace the varying scope of technology contributions, observing patterns from individual-level processes to systematic integration (Component 3). Throughout this process of applying the methodology shown in Figure 1., the conceptual framework is refined by moving between empirical patterns and theoretical foundations. The following subsections (2.1-2.4) present the thematic evidence base from which the components of the conceptual framework are derived. Section 3 then presents the resulting framework intended to guide digital technology design for social wellbeing in the built environment.

### 2.1 Social Wellbeing as a Conceptual Domain

Social wellbeing has been defined in various ways across disciplines, but a common thread is the emphasis on relational quality, community engagement and equitable participation. A foundational definition of social wellbeing (Keyes, 1998) is an individual's appraisal of their circumstances and functioning in society, comprising five dimensions: social integration, social acceptance, social contribution, social actualization and social coherence. This view makes social wellbeing especially relevant to studying technology in built environments, where interventions may affect some aspects of social life while leaving others unchanged. In the built environment literature, social wellbeing is closely linked to place and spatial design. Public spaces primarily serve social life, and design quality directly shapes opportunities for social interaction, gathering and civic participation (Gehl, 2013). The concept of third places (informal community anchors such as cafes, parks and community centres) (Maran, 2023) shows how spatial design supports social cohesion by providing settings for meeting outside home and work. Research on place attachment connects spatial experience to social belonging, with stronger attachment correlating with greater social connection and community engagement (Kyttä et al., 2016; Sánchez de Francisco et al., 2023; Scannell and Gifford, 2010; Shao et al., 2022). These foundations together with the synthesis of the literature presented in subsections 2.2 - 2.5 establish that social wellbeing is multidimensional and spatially embedded. Therefore, understanding how technology in built environments may strengthen or disrupt these spatial foundations requires a conceptual framework that accounts for both. The identified dimensions and subdimensions of the social wellbeing through the synthesis of the literature is illustrated in Figure 2.

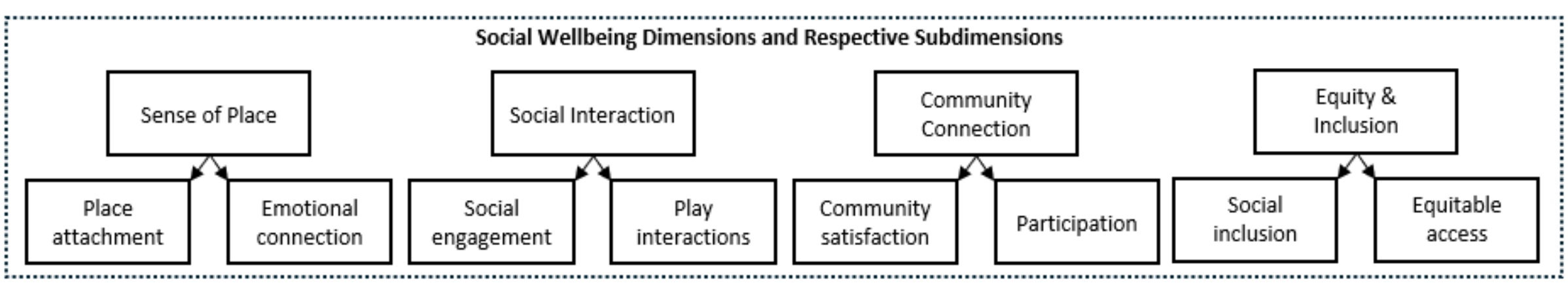

Figure 2: Dimensions and subdimensions of Social Wellbeing

## 2.2 Social Interaction

The empirical evidence indicates that technology facilitates social interaction in built environments through two distinct mechanisms with different implications for design. The first is spatial anchoring in which digital technology creates or enhances physical gathering points that draw people into shared spaces. Smart street furniture (Marshall et al., 2024), interactive urban installations (Nikolic and Yang, 2020), and technology-enhanced public amenities (Guo et al., 2024; Hatem et al., 2024) all function as social catalysts by giving people reasons to stop, linger and encounter others. These effects extend beyond direct technology users because non-users also experience increased social contact simply by being present in the activated space (Marshall et al., 2024). This suggests that well-placed technology can reshape the social dynamics of an entire setting.

The second mechanism is shared experience in which technology overlays digital content onto physical spaces in ways that create common reference points for social exchange. Interactive playgrounds (Van Delden et al., 2017a) improved social play patterns among children and caregivers with benefits extending beyond direct technology use. Interactive installations in community spaces (Back et al., 2018a) created opportunities for engagement which suggests that shared technology experiences can bridge rather than reinforce social boundaries. In healthcare settings, technology extended social connections beyond care facilities and help patients and elderly residents maintain community ties (Martin et al., 2024; Pilosof et al., 2023, 2021).

These findings establish social interaction as a dimension of technology-mediated social wellbeing. However, most of the reviewed studies measure immediate interaction frequency rather than the quality or sustainability of social connections over time and the distinction between technology as spatial anchor and technology as a shared experience has not been explicitly theorised despite its implications for design.

### 2.3 Community Connection

Research shows that digital technology supports processes through which individuals contribute to and identify with their communities. This occurs in three ways. The first is collective agency in which technology provides communities with tools to understand and shape their environments. Community based participatory research approaches (Rainisio et al., 2024) and context-sensitive digital urban planning tools (Kyttä et al., 2016) both show that when communities have data and decision-making capacity, the resulting empowerment strengthens individual and collective wellbeing. The second is shared meaning-making, in which technology-mediated environments create contexts for collective interpretation and dialogue. Experiential designed environments (Sharji et al., 2022) and digitally mediated public spaces (Sánchez de Francisco et al., 2023) both contributed to community satisfaction and social cohesion through processes of shared interpretation. The third is lowered barriers, where mobile applications and interactive installations (Papageorgiou et al., 2024; Zhang et al., 2023) increased community engagement by making participation more accessible and immediate.

A consistent pattern across these studies is the distinction between passive technology use and active community engagement. The social wellbeing benefits are strongest when technology supports genuine participation rather than passive consumption, suggesting that community connection functions as a distinct dimension of social wellbeing.

### 2.4 Sense of Place

The literature points the ways in which technology shapes social wellbeing by altering people's experiential relationship with the places they inhabit. The emotional bond between individuals and their environments operates bidirectionally with social wellbeing such as the stronger place attachment enhances social engagement while positive social experiences deepen one's bond with a place. Digital tools for urban happiness assessment (Kyttä et al., 2016) found links between place satisfaction and social connectivity, experiential designed environments (Sharji et al., 2022) enhanced visitor's sense of place through multimedia engagement which in turn enhanced social connection, and technology-enhanced user engagement (Sánchez de Francisco et al., 2023) strengthened place attachment and community belonging. A study of sensory awareness in built environments (Biggs, 2015) found that technology-mediated aesthetic experiences contributed to deeper place engagement and, through that, to social connection among users sharing those experiences. These findings align with the tripartite framework of place attachment (Scannell and Gifford, 2010), which identifies person, process and place dimensions.

The analytical significance of this literature is that technology may influence social wellbeing not only through direct social mechanisms but also through indirect experiential pathways. Sense of place functions both as a distinct outcome and as a mediator that amplifies other dimensions and it is largely overlooked in studies focused on social interaction metrics. This means that for framework design, sense of place must be assessed both as an outcome and as a mechanism influencing social interaction, community connection and inclusion.

### 2.5 Equity and Inclusion

The studies reviewed above document positive social outcomes, but only a few studies ask who benefits and who is excluded. This is a question with direct implications for whether technology-mediated social wellbeing is genuinely achieved or merely redistributed. An examination of both users and non-users of smart street furniture (Marshall et al., 2024) revealed concerns about equitable access, with the potential for digital divides to deepen rather than bridge social exclusion. Design choices play a decisive role because a study of interactive community spaces (Back et al., 2018a) found that age, cultural background and digital familiarity all influenced engagement patterns, while studies on smart home technologies (Geng et al., 2022; Tewell et al., 2019) found that IoT-based systems could support vulnerable populations but simultaneously excluded those lacking digital skills or economic resources. This pattern aligns with broader frameworks positioning equity as essential for community flourishing (Berkman et al., 2000; Keyes, 1998), where social acceptance is directly undermined when technology creates visible distinctions between those who can and cannot participate in digitally mediated social life.

These findings suggest that equity and inclusion function both as a dimension to be addressed in its own and as an enabling condition that determines whether benefits in other dimensions reach their intended breadth.

## 3. Conceptual Framework for Designing Technology for Social Wellbeing

The synthesis of the literature following the method presented in Section 2 resulted in a conceptual framework with three components where each component addresses an aspect of the technology design challenge. Together, they provide a structure for designing digital technologies that target social wellbeing in built environments. The first component identifies the dimensions and subdimensions of social wellbeing from the empirical literature and we try to establish what designers could address. The second is the technology contribution matrix where we classify the types of contributions technology could make which guide designers in how to orient their interventions when it comes to outcome or measurements. The third is the scope which maps the level at which technology could support social wellbeing started from individual processes to systemic integration and could help designers determine at what level to aim. Figure 3. provides an overview of the conceptual framework and its components. The rationale of the three components within the conceptual framework is that design guidance requires clarity about the target construct (what to address), the contribution types and measures (how to contribute and measure), and the scale of impact (at what level to aim), and no single component could address all three.

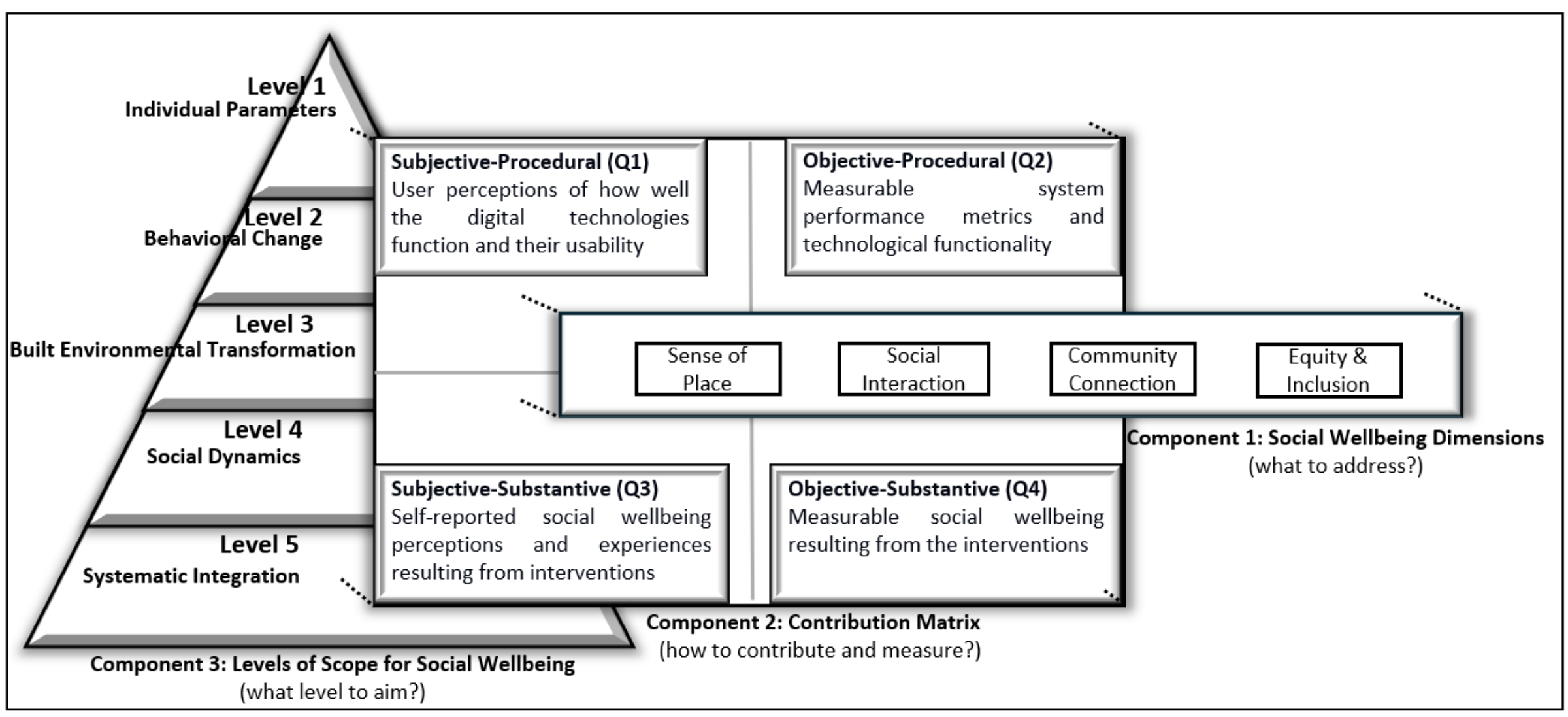

Figure 3: Conceptual Framework for Designing Technology for Social Wellbeing in Built Environments

### 3.1 Component 1: Dimensions and Subdimensions of Social Wellbeing

We identify from the literature that the social wellbeing in technology enabled built environments as a multidimensional construct with four interrelated dimensions (sense of place, social interaction, community connection and equity and inclusion) where each comprises two subdimensions (details give in Table 1.). In section 2, we reviewed these subdimensions as they appear across the empirical literature and in this subsection, we organize and described them into a structured analytical model that identifies the core areas technology designers could address to support social wellbeing comprehensively. The identification of these dimensions draws on both the empirical patterns observed across the reviewed studies and the multidimensional account of social wellbeing (Keyes, 1998). The specific four-dimension and it's subdimensions (shown in Figure 2. above and described in Table 1. below.) represent our conceptual contribution rather than a direct replication of any single source.

Table 1: Dimensions and Subdimensions of Social Wellbeing in Technology Mediated Built Environments

| Dimension | Subdimension | Description |
|---|---|---|
| **Sense of Place** | **Place Attachment** | The emotional and experiential bond between individuals and specific locations, promoted through technology-mediated environmental engagement. |
| | **Emotional Connection** | The affective responses and psychological ties individuals develop with built environments through technology-mediated experiences. |
| **Social Interaction** | **Social Engagement** | The quality, frequency, and nature of interpersonal contacts facilitated by technology in built environments. |
| | **Play Interactions** | Shared physical and playful experiences facilitated by interactive technologies that promote social bonding and cooperative behaviour. |
| **Community Connection** | **Community Satisfaction** | The degree to which individuals feel content with their community and its functioning, including collective identity and neighbourhood quality. |
| | **Participation** | Active engagement in community processes, collective decision-making, and shared activities facilitated by technology. |
| **Equity and Inclusion** | **Social Inclusion** | The extent to which technology-mediated built environments accommodate diverse populations and ensure access to social resources. |
| | **Equitable Access** | Fair distribution of technology-mediated social benefits and resources across population groups, addressing digital divides and structural inequalities. |

Furthermore, these identified dimensions do not operate in isolation. The literature shows significant interdependencies among them. For instance, the social interaction serves as an enabler such as the increased interpersonal contact in built environments is associated with improved community connection (Colangelo et al., 2023; Rainisio et al., 2024), greater participation in collective activities (Guo et al., 2024; Hatem et al., 2024; Nikolic and Yang, 2020), and better perceptions of inclusion (Marshall et al., 2024). Similarly, sense of place reinforces community connection by fostering emotional bonds with environments that motivate collective engagement (Rainisio et al., 2024; Sánchez de Francisco et al., 2023). Moreover, literature showed that when technology-mediated environments are inclusive, they expand the base of individuals who benefit from social interaction, community connection, and sense of place (Back et al., 2018b; Marshall et al., 2024). These interdependencies have important implications for design because digital technology targeting one dimension of social wellbeing may produce cascading effects across others and evaluation must account for these links rather than treating each dimension in isolation.

### 3.2 Component 2: Digital Technology Contribution Matrix

The second component provides a matrix for understanding the types of contributions that digital technology can make to social wellbeing and ways to measure it. We identify two complementary dimensions i.e., the nature of outcomes (procedural vs. substantive) and the type of evidence (subjective vs. objective) based on the analysis of how empirical studies measure and report technology contributions. These dimensions define a four-quadrant matrix that captures how digital technology can contribute to social wellbeing. Figure 4. provides a visual overview of the matrix as well as presents the details of each quadrant within the matrix.

| | Subjective | Objective |
|---|---|---|
| Procedural | **Subjective-Procedural (Q1)**<br>User perceptions of how well the digital technologies function and their usability<br>**Contribution to Social Wellbeing:**<br>*Digital technology contribute to social wellbeing by creating positive user experiences and removing barriers to technology adoption, facilitating sustained engagement.* | **Objective-Procedural (Q2)**<br>Measurable system performance metrics and technological functionality<br>**Contribution to Social Wellbeing:**<br>*Digital technology contribute to social wellbeing by ensuring consistent, accurate, and timely technological function, creating a reliable foundation for social wellbeing* |
| Substantive | **Subjective-Substantive (Q3)**<br>Self-reported social wellbeing perceptions and experiences resulting from interventions<br>**Contribution to Social Wellbeing:**<br>*Digital Technology contribute to social wellbeing by enhancing subjective experiences w.r.t to each subdimension of social wellbeing.* | **Objective-Substantive (Q4)**<br>Measurable social wellbeing resulting from the interventions<br>**Contribution to Social Wellbeing:**<br>*Digital Technology contribute to social wellbeing by creating measurable improvements in social wellbeing.* |

Figure 4: Digital Technology Contribution Matrix for Social Wellbeing

The distinction between procedural and substantive outcomes reflects a key insight that technology affects social wellbeing through both the processes it enables and the tangible social changes it produces (Greenhalgh et al., 2017; Weiser and Brown, 1997). Procedural outcomes concern technological functions that create conditions for social wellbeing, such as interaction design quality, usability and reliability. Substantive outcomes concern actual social changes, such as increased interaction, stronger community ties or enhanced inclusion. This distinction matters because strong procedural performance does not automatically produce substantive social benefits (Venkatesh et al., 2003). A well-functioning interactive installation may fail to generate meaningful social encounters if it is poorly located, culturally irrelevant, or designed without attention to social dynamics. Design must attend to both the process and its social consequences.

The distinction between subjective and objective measurement reflects social wellbeing's dual nature as both are experienced and an observable phenomenon. Subjective measures capture how individuals perceive their social circumstances (belonging, community satisfaction, perceived inclusion). Objective measures capture observable aspects (interaction frequencies, participation rates, demographic diversity of engagement). Research shows that integrating both types is essential for comprehensive wellbeing-oriented design (Keyes, 1998; Neff and Olsen, 2007), as they may diverge notably. A community may show high interaction frequency (objective) while residents report low satisfaction with interaction quality (subjective), or vice versa. Capturing only one type risks a misleading picture of social wellbeing.

The matrix could allow mapping a planned intervention across all four quadrants and helps identify which aspects of social wellbeing are being targeted and which are neglected. For example, a smart public installation might aim primarily at procedural-subjective outcomes (user satisfaction with social features) while overlooking substantive-objective outcomes (measurable changes in community participation rates). The matrix makes such gaps visible, prompting designers to consider a fuller range of contributions. Additionally, the matrix could enable comparison across interventions because different technology types tend to concentrate their contributions in different quadrants and understanding these patterns could help designers learn from precedents and position their design strategically.

### 3.3 Component 3: Scope Levels

The third component maps the scope to which digital technology can contribute to social wellbeing by identifying levels of contribution (shown in Figure 5.).

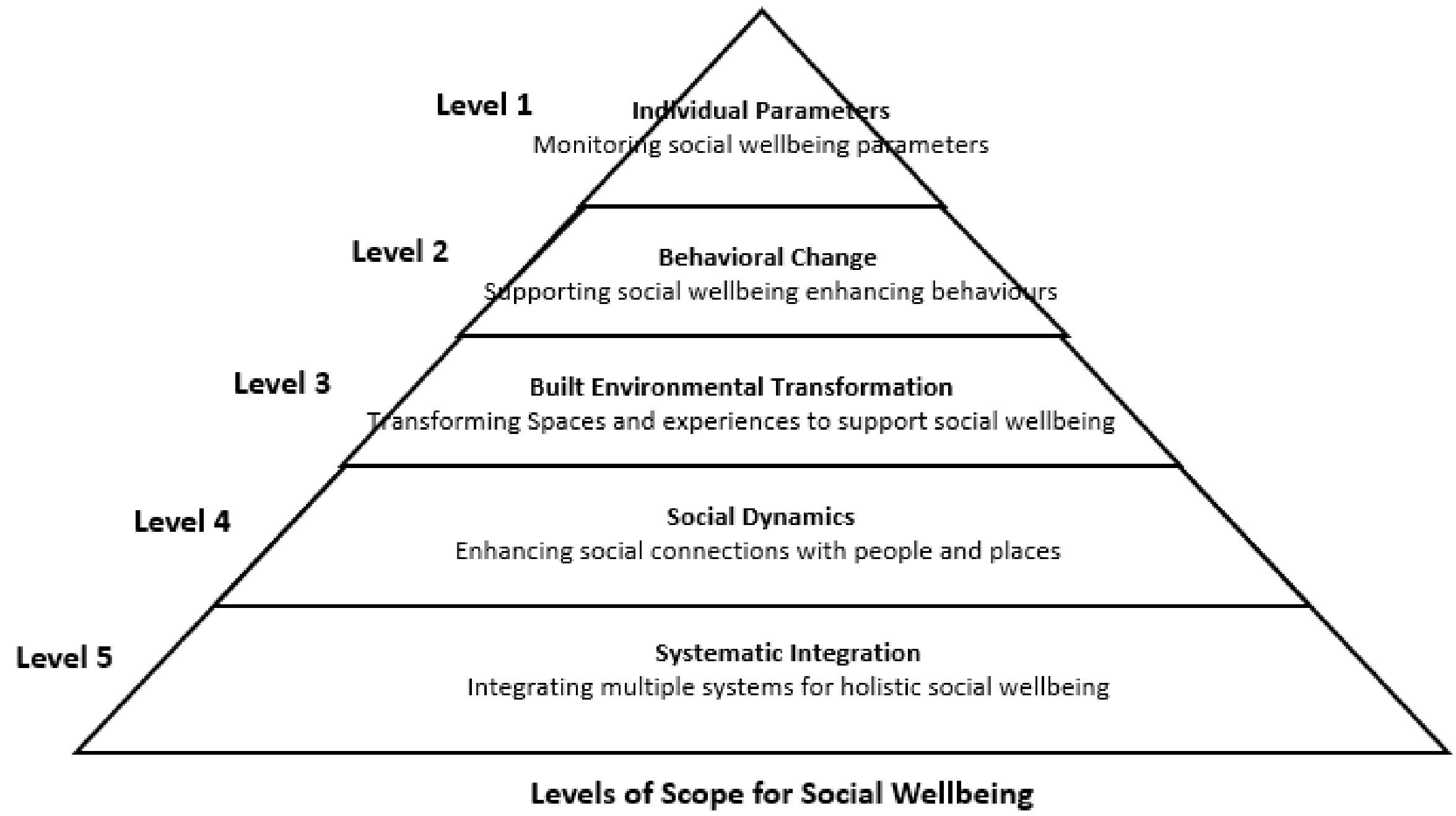

Figure 5: Levels of Scope for Digital Technology to Social Wellbeing

The levels vary from foundational individual-level processes to systematic integration. The layered structure reflects a key finding that the digital technology's social wellbeing contributions vary in scope. Each level represents a scope rather than a step in a sequence. At the narrowest scope, individual awareness (Level 1) addresses how technology makes social patterns, needs, or opportunities visible to individuals. Behavioural activation (Level 2) encompasses a wider scope in which technology prompts or supports changes in social behaviour. Built environment transformation (Level 3) extends further by capturing how technology reshapes the physical or digital conditions under which social wellbeing occurs. Social dynamics (Level 4) encompasses collective effects on social cohesion, participation and belonging. Systemic integration (Level 5) represents the broadest scope, where multiple technological and social systems are coordinated for overall social wellbeing outcomes. This structure aligns with socio-ecological perspectives (Stokols, 1996) and the findings (Yang and Kim, 2024) that technology in built environments affects physical, psychological and social dimensions of human experience simultaneously. These levels help designers understand the scope of social wellbeing contribution they are aiming for and what capabilities must be in place. The Figure 5. above illustrates this visually, and Table 2. give details of each level with its description, social wellbeing focuses and contribution emphasis. In the column named "Contribution Matrix Emphasis", we try to connect each scope level to the technology contribution matrix (Section 3.2), indicating which types of outcomes and evidence become most important as the scope of impact broadens.

Table 2: Levels of Scope of Digital Technology to Social Wellbeing

| Level of Scope | Description | Social Wellbeing Focus | Contribution Matrix Emphasis |
| --- | --- | --- | --- |
| **Level 1: Individual Social Awareness** | Monitoring and making visible individual social parameters, behaviours, and environmental conditions relevant to social wellbeing | Establishing baseline awareness of one's social environment, community context, and patterns of social engagement with people and place | Primarily Q2 (objective-procedural) and Q4 (objective-substantive): data collection quality and awareness outcomes |
| **Level 2: Social Behavioural Activation** | Promoting and supporting behaviours that enhance social engagement | Activating social behaviours such as encouraging interaction, facilitating participation, prompting community engagement | Balanced across all quadrants: behavioural triggers (Q1, Q2) producing observable and perceived social changes (Q3, Q4) |
| **Level 3: Built Environment Transformation** | Transforming physical and digital spaces to create environments that actively support social wellbeing | Reshaping environments to enhance social interaction opportunity, sense of place, and community identity | Strong emphasis on Q3 (subjective-substantive): sense of place, community identity, and experiential quality |
| **Level 4: Social Dynamics** | Enhancing social connections, community cohesion, and collective dynamics through technology-mediated processes | Strengthening community ties, fostering social inclusion, enhancing collective participation | Strong emphasis on Q3 and Q4 as both perceived and measurable community level social outcomes |
| **Level 5: Systemic Social Integration** | Integrating multiple systems, stakeholders, and domains to achieve holistic and sustainable social wellbeing | Achieving coordinated, equitable, and sustainable social wellbeing across interconnected systems and diverse populations | Comprehensive across all quadrants, with particular attention to equity, sustainability, and systemic coordination |

The levels represent nested scopes rather than a linear sequence and digital technologies could target multiple levels simultaneously, and effects at one level can reinforce or reshape others. Improvements in social dynamics (Level 4) could strengthen individual awareness and behavioural activation (Levels 1–2) by creating more socially supportive environments in which people are more adapted to social cues and more willing to engage. Built environment transformations (Level 3) may attract new user groups whose presence alters neighbourhood dynamics (Level 4), which in turn shifts individual experiences (Levels 1–2). These interactions mean that the effective digital technologies are often those designed with multiple levels in view, using narrower-scope contributions to support broader ones and vice versa. This is therefore best understood as a conceptual map of impact scope that helps designers determine which level of contribution their intervention aims to achieve and how different scopes reinforce one another.

## 4. Framework Integration and Operationalization

This section integrates the framework components presented in Sections 3 to practical use by providing guidance on contribution measurement tool and design for each subdimension of social wellbeing (see Table 3.). Additionally, the operationalization protocol of the conceptual framework is explained below and illustrated in Figure 6. The measurements column in the Table 3. identifies how each subdimension can be assessed, specifying adapted indicators and data collection approaches. The design guidance column translates the empirical mechanisms identified in Section 2 into design recommendations for each subdimension, directing practitioners toward specific design strategies.

Table 3: Social Wellbeing Measurement and Design Guidance for Digital Technology in Built Environments

| Social Wellbeing Dimension | Social Wellbeing Subdimension | Measurements | Design Guidance |
|---|---|---|---|
| **Sense of Place** | **Place Attachment** | Adapt place attachment items to capture technology mediated place experiences; measure attachment to both physical and digitally augmented aspects of place. | Use technology to reinforce existing place identity rather than overwrite it; embed digital features that respond to local history, culture, or spatial character; use technology to make the history, cultural layers, or ecological qualities of a place visible to inhabitants. |
| | **Emotional Connection** | Distinguish emotional responses to the technology itself from emotional responses to the environment as mediated by technology; include items capturing sensory enrichment through digital augmentation. Capture sensory enrichment and aesthetic experience through digital augmentation. | Prioritize technology that enhances built environmental qualities; ensure interventions deliver experiential value beyond functional utility. |
| **Social Interaction** | **Social Engagement** | Supplement self-report with sensor-based interaction detection (e.g., proximity sensing); distinguish technology-facilitated encounters from incidental co-presence; capture both online and co-located interactions. Measure both interaction frequency and reported interaction quality over time. | Apply the spatial anchoring mechanism by positioning technology to create natural gathering points; apply the shared experience mechanism by overlaying digital content that gives strangers a common reference point for conversation. |
| | **Play Interactions** | Assess whether technology supports or replaces spontaneous physical play; measure cross-generational engagement patterns; capture cooperative versus parallel play behaviours in interactive installations. Capture whether play interactions extend into ongoing social relationships beyond the technology encounter. | Ensure interactive installations require or reward cooperation rather than individual use; design for variable skill levels so that adults, children, and older adults can participate together; balance digital novelty with sustained playability to counter habituation effects. |
| **Community Connection** | **Community Satisfaction** | Include items on technology's contribution to perceived community quality; measure whether digital infrastructure enhances or fragments community identity; assess satisfaction with technology-mediated community services. | Support the collective agency by giving communities visible, shared data about their neighbourhood; ensure digital platforms represent the community's character rather than imposing generic interfaces. |
| | **Participation** | Distinguish passive technology consumption from active community engagement; measure co-creation and collective decision-making through digital platforms; assess whether participation translates to perceived empowerment. | Provide platforms for co-creation and collective decision-making, not just information delivery; lower participation barriers through accessible interfaces and multilingual support; ensure that digital participation translates to perceived empowerment and tangible community change. |
| **Equity & Inclusion** | **Social Inclusion** | Assess both digital and physical accessibility barriers; measure demographic diversity of technology users versus the broader population; evaluate whether technology creates new forms of exclusion. | Apply universal design principles to all technology interfaces; provide non-digital alternatives for core social functions; conduct user testing with underrepresented groups during design rather than after deployment. |
| | **Equitable Access** | Measure benefit distribution across socioeconomic, age, and ability groups; assess cost and literacy barriers to technology use; evaluate representation of diverse groups in technology design and governance processes. | Minimize dependence on personal devices or data plans for core social functions; ensure technology-mediated benefits are available through multiple access channels; involve diverse population groups in design governance to ensure that equity considerations are integrated into technology from the outset (to ensure fair, needs-based access and reduce structural barriers). |

The Table 3. must be used alongside the digital technology contribution matrix (Figure 4., Section 3.2) and scope levels (Table 2., Section 3.3) and could help connect specific design choices to quadrants of the contribution matrix and levels of scope. To explain this further, we proposed a five-step protocol (illustrated in Figure 6.), that could guide practitioners and researchers to apply the conceptual framework to a specific digital technology design for social wellbeing in the built environment.

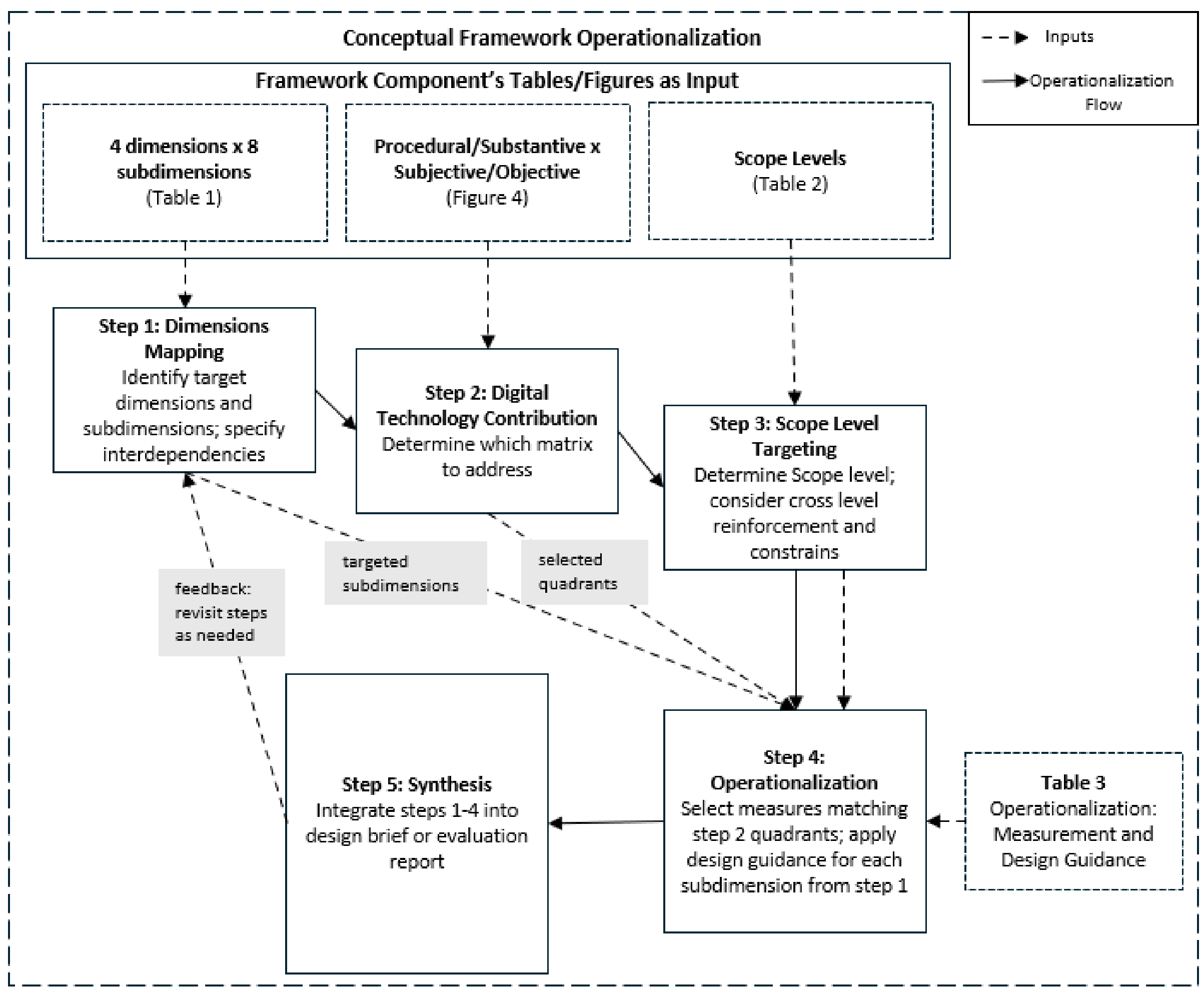

Figure 6: Operationalization Protocol for Designing *Technology* for Social Wellbeing

Step 1: Identify which dimensions of social wellbeing the digital technology could target using the dimensions and subdimensions model (Table 1.). Additionally, specify the interdependencies among the dimensions or subdimensions most relevant to the context. Step 2: Determine using the digital technology contribution matrix (Figure 4.) which quadrants the digital technology could address. Identify any quadrant gaps where contributions may be missing and assess whether addressing them would strengthen the design. Step 3: Determine which scope levels the intervention aims to reach by using the scope levels (Table 2.). Consider which levels the intervention addresses simultaneously and how effects at one level may reinforce or constrain others. Step 4: Use Table 3. to translate each targeted subdimensions of social wellbeing identified in Step 1 into concrete measurement and design decisions. Select the measures that correspond to the matrix quadrants identified in Step 2. For example, if need for subjective-substantive evidence (Q3) is identified for place attachment, Table 3. directs adapting place attachment items to capture technology-mediated place experiences; if objective-procedural evidence (Q2) is needed for social engagement, it directs supplementing self-report with sensor-based interaction detection. Additionally, the design guidance column specifies how to design for each subdimension. For example, social engagement directs applying spatial anchoring by positioning technology to create natural gathering points and overlaying digital content that gives strangers a common reference point for conversation, while for equity and inclusion, it directs applying universal design principles and providing non-digital alternatives for core social functions. Use Table 3. if multiple subdimensions are targeted to check that measurement and design choices are coherent across them. Step 5: Integrate findings from Steps 1–4 to produce a comprehensive design brief or evaluation report. The synthesis could explicitly address which dimensions and subdimensions are targeted and how their interdependencies shape the digital technology design; which contribution matrix quadrants provide evidence and whether the balance between procedural and substantive, subjective and objective contributions are adequate; and how the measurement and design choices from Table 3. work together across the targeted subdimensions to form a coherent intervention rather than a collection of isolated features. Revisit the steps 1-4 if needed.

## 5. Discussion

This chapter proposed a conceptual framework for designing digital technology that supports social wellbeing in built environments. Social wellbeing is the quality of relationships, community engagement and sense of belonging that people experience in everyday life (Keyes, 1998). This is shaped in various ways by the physical and digital environments people inhabit. Similarly, people feel connected to the places and people around them and are influenced by how public spaces are designed, what technologies are embedded in them, and whether those technologies invite participation. The consequences of neglecting social dimension are well documented, such as the social isolation and loneliness are associated with elevated mortality risk and poor physical and mental health outcomes (Hodgson et al., 2020; Holt-Lunstad, 2024; Wang et al., 2023), and the World Health Organization (WHO) has called for social health to be treated with the same urgency as physical and mental health (Garcia et al., 2025). Although research recognizes that digital technologies could facilitate social interaction, strengthen community ties, and foster place attachment in built environments (Back et al., 2018b; Sánchez de Francisco et al., 2023; Sharji et al., 2022; Van Delden et al., 2017b), the designers of such technologies have had little structured guidance for addressing social wellbeing from all perspectives rather than in isolated fragments.

The conceptual framework proposed in this chapter builds upon established theoretical foundations and extends them in ways that existing applied frameworks have not. Keye's (Keyes, 1998) conceptualization of social wellbeing as encompassing social integration, contribution, coherence, actualization and acceptance provided the starting point. However, as a general psychological construct, it does not account for how physical environments or digital technologies shape these experiences. Built environment assessment frameworks (Akbarinejad et al., 2023; Makaremi et al., 2025) do acknowledge the role of the built environment, but treat social outcomes as aggregated indicators alongside physical, mental and environmental ones, without specifying what social wellbeing consists of or how technology contributes to it. The social wellbeing dimensions (component 1 of our conceptual framework) addresses both limitations by grounding four dimensions and subdimensions in empirical evidence of what changes when digital technology is introduced into built environment. Additionally, socio-ecological theory (Stokols, 1996) has long argued that outcomes are shaped by nested levels from individual to societal and the scope levels (component 3 of our conceptual framework) reflects this thinking. However, existing smart city frameworks that draw on similar logic tend to reduce the social dimension to citizen satisfaction or participation rates (Freestone and Favaro, 2022)without distinguishing how deep a technology's contribution reaches. Our framework maps five nested levels from individual awareness to systemic integration and offers the specificity that these broader frameworks lack. The contribution matrix (component 2 of our conceptual framework) draws on a distinction present in both technology assessment (Greenhalgh et al., 2017) and user experience research (Hassenzahl and Tractinsky, 2006) that technology affects people through both the processes it enables and the outcomes it produces. Digital health technology assessment frameworks (Segur-Ferrer et al., 2024; Vis et al., 2020) apply a similar evaluative logic, but focus on clinical effectiveness, safety and cost, with limited room for relational and community-level outcomes. The digital wellbeing literature (Büchi, 2024) has expanded this focus to individual experiences of technology use, but not to how technology mediates social connections within physical spaces. Our conceptual framework addresses this by distinguishing procedural from substantive outcomes and subjective from objective measurement by providing an evaluation structure suited to social rather than clinical or individual outcomes. In each case, the gap is not that social outcomes are absent from existing work but that they lack the specificity, measurement structure and scope that a dedicated framework could provide.

The operationalization process described in Section 4 showed that these components translate into practical measurements and design decisions. However, the value of any framework depends not only on its conceptual coherence, but on whether it is useful to the people who might apply it. The following subsection considers what the framework offers to different entities that design and engage with digital technology for social wellbeing in built environments.

### 5.1 Conceptual Framework Implications

The framework enables studies for researchers that move beyond isolated social indicators toward a more complete account of how digital technology shapes social wellbeing in built environments. Rather than measuring only whether technology increases social interaction or community satisfaction, researchers could now design studies that ask what type of social outcome is affected, whether the evidence is procedural or substantive, subjective or objective, and at what level the contribution operates, from individual through to systemic. These connections across dimensions, contribution types and scope are what allow research to capture the full picture rather than fragments of it. This integrated logic is equally relevant for health and technology assessment, where social wellbeing remains underrepresented. Existing built environment assessment frameworks (Makaremi et al., 2025; Tan et al., 2025) treat social outcomes as secondary indicators, and digital health technology assessment frameworks focus on clinical efficacy, safety and cost (Segur-Ferrer et al., 2024; Vis et al., 2020), with limited attention to social and relational outcomes. The framework offers these evaluation traditions a way to assess social wellbeing with the same structure and specificity they currently apply to other health dimensions.

For practitioners and policymakers, the framework's value lies in connecting design decisions to evaluation criteria within a single coherent process. Rather than treating social wellbeing as an add-on, the conceptual framework operationalization protocol (Section 4) allows urban planners and technology designers to identify which social outcomes to target, determine what evidence would demonstrate success, and set realistic expectations about how deep the intervention's contribution could reach. At the policy level, the same integrated structure could inform funding criteria, procurement requirements and post-deployment evaluation by providing a shared language for specifying what social benefits technology projects could deliver and how those benefits could be assessed.

### 5.2 Limitations

This research work advances the current understanding of social wellbeing in digital technology mediated built environments. However, it has limitations that point to areas for further development. The framework is derived from a focused integrative review (Jabareen, 2009) and the body of empirical evidence on technology-mediated social wellbeing remains limited in both volume and geographic scope. Therefore, the framework's dimensions, measurement approaches and scope levels may require refinement or expansion to accommodate different cultural conceptions of social wellbeing and different patterns of technology use. What constitutes meaningful social interaction, community participation, or equitable access varies across cultures, socioeconomic conditions, and governance structures. A digital technology designed for social engagement in one setting may not produce the same outcomes in another where social norms, digital competencies, language, and patterns of built environment use differ fundamentally. This framework addresses contextual adaptation only indirectly, and more explicit integration of context sensitivity into its dimensions, levels, and application protocol represents a direction for future development.

Additionally, the framework is conceptual and remains to be validated through application to specific design projects. The conceptual framework is a synthesis of the findings of empirical studies and integration of those findings into a component structure involves interpretive choices that may not capture all relevant dimensions or relationships. Empirical application in real design contexts would test the framework's practical utility and reveal dimensions that may need refinement or addition. Last but not the least, the social wellbeing dimensions and their subdimensions are presented as relatively distinct categories, but in practice they overlap significantly and may not be cleanly separable in empirical measurement. The framework acknowledges these interdependencies conceptually but does not yet provide specific analytical guidance for handling measurement overlaps, such as mediating or moderating relationships between subdimensions. Future methodological work could develop structural models that test the relationships among subdimensions and provide more precise tools for disentangling their effects.

## 6. Conclusion

This chapter addressed the lack of structured guidance for designing digital technology that supports social wellbeing in built environments. Social wellbeing is a recognized component of the multidimensional model of health and wellbeing, and it is an important pathway through which digital technologies within built environments can influence broader health outcomes. However, social dimension has received less attention than its physical, mental counterparts. Therefore, we have developed a conceptual framework by drawing on empirical studies of technology interventions in built environments for social wellbeing. The framework comprises three integrated components i.e., a social wellbeing dimensions model that identifies four dimensions and relevant subdimensions of social wellbeing, a digital technology contribution matrix that distinguishes procedural from substantive outcomes and subjective from objective evidence, and scope levels that map five nested levels at which technology could contribute to social wellbeing.

The conceptual framework's contribution is not the individual components themselves but their integration into a coherent structure that connects what to design for, how to contribute and measure it, and what level the contribution reaches. Although existing approaches address parts of this challenge such as the built environment assessment frameworks acknowledge social outcomes, they treat them as secondary indicators. Additionally, digital wellbeing research centres on individual technology use rather than social connections in physical spaces, and health technology assessment frameworks provide evaluative rigour but focus on clinical rather than relational outcomes. The conceptual framework we propose bridges these partial perspectives by offering a structure in which social wellbeing dimensions, measurement logic, and scope levels work together. This integration allows researchers to design studies that capture social wellbeing in its full breadth, practitioners to link design decisions to evaluation criteria, policymakers to set clear expectations about social benefits, and health and technology personnel to address social outcomes with the same specificity currently applied to other health and wellbeing dimensions and supports a promotive and preventive approach to health.

The conceptual framework remains conceptual and awaits empirical validation through application in specific design and assessment contexts. Future work could test the framework across culturally and socioeconomically diverse settings, where different conceptions of social wellbeing and different patterns of technology use may require refinement of the dimensions or scope levels. Moreover, methodological development is also needed, particularly structural models that examine the relationships and interdependence among subdimensions of social wellbeing and provide guidance for handling their empirical overlap. Meanwhile, more explicit integration of contextual factors, such as governance structures, cultural norms and socioeconomic conditions, would strengthen the framework's adaptability. We emphasise that to truly promote health and wellbeing, it is crucial to develop equity-oriented digital solutions for built environments that not only address individual needs but are also systematically integrated into planning and decision-making processes at all levels. The need for structured approaches to social wellbeing will only grow as built environments become increasingly mediated by digital technology. This conceptual framework is a step toward ensuring that when we design digital technology for built environments, social wellbeing is addressed with the same rigour as other dimensions of health and wellbeing. As a next step in this research, we plan to use the framework to guide the design and evaluation of a digital technology intervention for social wellbeing in a built environment. This will provide an empirical test of the framework's practical utility and reveal which components require refinement when applied to a specific design context.

**Acknowledgments:** This research has been funded by the SWELL project (https://www.ntnu.edu/sustainability/swell) and NTNU strategic area Sustainability (https://www.ntnu.edu/sustainability).